Title: Protein-protein interactions enhance the thermal resilience of SpyRing enzymes: a molecular dynamic simulation study

Running title: enhancing thermal resilience by adjusting SpyRing enzyme interface

Qi Gao and Dengming Ming*

College of Biotechnology and Pharmaceutical Engineering, Nanjing Tech University, 30 South Puzhu Road, Jiangbei New District, Nanjing City, Jiangsu, 211816, PR China

* Corresponding author. Biotech Building Room B1-404, College of Biotechnology and Pharmaceutical Engineering, Nanjing Tech University, 30 South Puzhu Road, Jiangbei New District, Nanjing City, Jiangsu, 211816, PR China.
E-mail address: dming@njtech.edu.cn.

# ABSTRACT

Recently a technique based on the interaction between adhesion proteins extracted from *Streptococcus pyogenes*, known as SpyRing, has been widely used to improve the thermal resilience of enzymes, the assembly of biostructures, cancer cell recognition and other fields. In SpyRing, the two termini of the target enzyme are respectively linked to the peptide SpyTag and its protein partner SpyCatcher. SpyTag spontaneously reacts with SpyCatcher to form an isopeptide bond, with which the target enzyme forms a close ring structure. It was believed that the covalent cyclization of protein skeleton caused by SpyRing reduces the conformational entropy of biological structure and improves its rigidity, thus improving the thermal resilience of the target enzyme. However, the effects of SpyTag/ SpyCatcher interaction with this enzyme are poorly understood, and their regulation of enzyme properties remains unclear. Here, for simplicity, we took the single domain enzyme lichenase from *Bacillus subtilis* 168 as an example, studied the interface interactions in the SpyRing system by molecular dynamics simulations, and examined the effects of the changes of electrostatic interaction and van der Waals interaction on the thermal resilience of target enzyme. The simulations showed that the interface between SpyTag/SpyCatcher and lichenase is different from that found by geometric matching method and highlighted key mutations that affect the intensity of interactions at the interface and might have effect on the thermal resilience of the enzyme. Our calculations provided new insights into the rational designs in the SpyRing.

# Introduction

Enzymes have been widely used in industry because of their superlative catalytic efficiency, stereoselectivity, mild reaction conditions and environmental friendliness, and more than a hundred enzymes have found large-scale industrial uses[1]. However, many promising enzymes developed in laboratory have encountered great challenges in industrial application, due to their poor stability at high processing temperatures[2], easy deactivation and difficult storage[3]. To meet the challenges, conventional methods, such as directed evolution, rational design and other structural or library based methods have been developed and gained remarkable success[4-6]. Since enzymes vary greatly in structure and function from one another, and the above-mentioned methods of enzyme modification can only be carried out on a case-by-case basis, lacking a unified guidance scheme[7, 8].

In the past few decades, protein cyclization technology has developed into as a generic approach to improve the thermal stability of enzymes. In protein cyclization, the N- and C-termini and other unfolded regions of a protein are covalently tethered so as to increase the rigidity of the protein, leading to higher entropy barrier when unfolding [9-11]. Various protein cyclization strategies, such as chemical synthesis[12], transpeptidation using sortase[13], intein-mediated splicing[14], have been reported to increase thermal resilience to a certain extent. Very recently, a novel cyclization approach called SpyRing, originally discovered in *Streptococcus pyogenes[15]*, has been reported using a SpyTag/SpyCatcher-mediated spontaneous isopeptide bond formation strategy, which can dramatically improve the thermal resilience[16, 17]. The basic mechanism is that when the Tag/Catcher genes are fused to the N- and C- termini of the enzyme, they spontaneously and rapidly react to form an irreversible iso-peptide bond between ASP of SpyTag and LYS of SpyCatcher, leading to the covalent cyclization of the enzyme. One of the advantage of this reaction is that it can be carried out under a variety of different conditions, including reduction or oxidation conditions and a fairly wide pH range[18, 19].

In the past few years, several research groups have reported to use the SpyRing technique to increase the thermal resilience of various enzymes. Schoene and colleagues showed that SpyRing cyclization leaded to >60ºC increases in thermal resilience of β-lactamase, which is much higher than conventional point mutation method can achieve[20]. Wang and colleagues applied SpyRing to cyclize lichenase from *Bacillus subtilis* 168 and raised the optimum temperature by 5ºC[21]. It is impressive that the cyclized enzyme still maintains 80% of its catalytic activity at 100°C, while the linear enzyme loses almost all of its catalytic activity at this temperature. Wang and colleagues studied the thermostability and organic solvent tolerance of cyclized L-phenylserine aldolase, and reported that cyclization increased the half-life of the enzyme by 8.3 times at 70°C and $T_{50}$ by 10.3°C[22]. Very recently, Zhou and colleagues compared the performances of linear- and cyclized-xylanase, which has potential sustainable commercial development value in green energy manufacturing industry. They found that although SpyRing did not significantly increase the optimal temperature of the enzyme, it did enhance thermal stability, ionic stability and resilience to aggregation and freeze-thaw treatment, without affecting its catalytic efficiency[23]. Si and colleagues used the cyclization technique to enhance the thermal resilient performance of a firefly luciferase (FLuc) from *Photinus pyralis*, a reporter enzyme that has many potential academical research and industrial applications[24]. The cyclized enzyme has the optimum temperature of 35°C, which was 10°C compared with the linear FLuc. In addition, they found the half-lives $t_{1/2}$ (at 45°C) and the melting temperature $T_m$ of the cyclized enzyme increased by 2.4 times and 16.7°C, respectively.

So far, SpyRing cyclization is primarily designed for monomeric enzymes. It was considered that the introduction of SpyCatcher/SpyTag at protein termini should avoid the catalytic sites of enzyme, thus improving the thermal stability of the enzyme without affecting their catalytic ability. It was considered that introduction of SpyRing should not hinder chaperone-assisted protein folding/refolding or the cofactor assisted protein refolding[25]. Application of SpyRing to multimeric enzyme systems required detailed consideration of the relative configuration of the SpyTag/SpyCatcher and the enzymes. In addition, in order to broaden the application scope of SpyRing, different mutations have been made to the system so as to achieve noval functions. Cao and colleagues introduced more than 10 negative-charge mutations in SpyTag/SpyCatcher, carrying a net charge of -21, and obtained a sensitive system whose cyclization reaction is tunable by environment PH[26]. Through reasonable design and directed evolution, Liu and colleagues reported that three mutations in SpyTag/SpyCatcher can lead to orthogonal reactivity of proteins with high sequence similarity, and achieved valuable characteristics such as high selectivity, inverse temperature dependence and orthogonal reactivity[27]. Keeble and colleagues obtained a series of mutants of SpyTag/SpyCatcher through a phage-display platform and rational design[28, 29]. Through the optimization of docking reaction and cyclization reaction, the obtained mutants greatly accelerated the reaction in order of magnitude, expanding the application of the system in detecting the interaction between enzymes and other biological structures. Very recently, Tian and colleagues introduced a special peptide (LysGlyLysGlyLysGly) into the C-terminus of SpyCatcher and three other Lys-Arg mutations around the binding domain. They found that the mutant SpyCatcher was more effectively attached to the glyoxyl-agarose support, making it highly binding to the SpyTag-fused protein[30]. When applying SpyRing technology, mutations were chosen only to optimize the cyclization reaction between SpyTag and SpyCatcher, in addition to requirement that the enzyme active site should not be blocked by SpyTag/SpyCatcher; Their potential effects on target enzymes through interface interaction were often ignored. In a recent study of SnoopRing cyclized luciferase, we found that cyclization introduced important interactions at the interface between the enzyme and Catcher/Tag[31]. A combination of molecular dynamic simulation and SDS-PAGE analysis revealed that the docking process of SnoopRing with luciferase was regulated by Lys22 in Tag and Lys557 in luciferase. The final compact structure involved the electrostatic interaction between Arg17 in Catcher and Asp520 in luciferase, hydrogen bond and van der Waals interactions. In this sense, we suspected that SpyRing-mediated cyclization should also introduce important interactions at the interface between the enzyme and the SpyTag/SpyCatcher structure. For simplicity, in this study we chose a single domain enzyme, lichenase[21], as an example to study the effect of cyclization introduced by SpyRing and its mutants on the performance of the target enzyme, especially its thermal stability. We used molecular dynamics simulations to determine the protein-protein interface (PPI) between the enzyme and SpyRing and mutants, which can be regarded as the result of restricted protein-protein docking. We then investigated the regulatory effect of SpyRing mutations on enzyme properties through PPI. Our calculations suggested the pathways involving the docking of SpyTag/SpyCatcher with the enzyme, highlighted key residues that regulate interactions between enzyme and SpyTag/SpyCatcher, and their effects on enzyme properties such as thermal resilience.

## Methods

*Molecular dynamics simulation of the cyclized lichanse structure*

In order to investigate the interaction between SpyTag/SpyCatcher and the enzyme as well as the effect of mutations at protein-protein interface on the thermal stability of the enzyme, we conducted a series of simulations of the cyclized system and those with a few designed mutants. The X-ray structure of β-1,3–1,4-glucanase (lichenase) from *Bacillus subtilis* 168 was taken from the protein data bank (PDB)[32] with entry code 3O5S[33], 2.2Å resolution. The start structure of the wild-type SpyTag/SpyCatcher was built using the homology modelling program MODELLER with the X-ray structure (PDB entry code 4MLI, 2.1Å resolution[19]) as a template, adding missing residues. The termini of the enzyme were linked to SpyRing with a six-residue GSGGSG linker as described in ref[21]. In the initial configuration, the SpyTag/ SpyCather was positioned not to contact the enzyme except that it was connected to it by a linker (Figure 1).

Simulations were performed by GROMACS[34] software. The gromos53a6 force field[35] was used to generate topology files of the original complex and the mutated complex. The SPCE water model was chosen so that the initial structure of each system was solvated in a cubic box with length of 112.12 Å. An appropriate amount of sodium counterion was added to neutralize each system so that it remained electroneutral. To minimize the energy, the system was optimized using the steepest descent minimization. Subsequently, the protein was subjected to a position restraint, after which each system was slowly heated to 333 K with a constant volume and equilibrated. Finally, a MD equilibration simulation of 100 ns was performed. Long range electrostatics with the cutoff distances of 1.0 nm and van der Waals were calculated using the particle mesh Ewald method (PME)[36]. The time step was set to 2 fs, and trajectories were collected at 10 ps intervals. The trajectories were analyzed using the gmx built-in programs -rms, -gyrate, -rmsf. Visualization programs VMD[37] and PyMOL[38] software were used to examine the molecular dynamics trajectories against the resulting graphical files of the simulations and to analyze their protein interfaces.

*Introductions of mutations to the cyclized lichenase*

A visualization of the simulation system obtained from the simulation showed that a stable interface was formed between wild-type type SpyTag/SpyCather and the enzyme, and different types of atomic interactions have been identified, including electrostatic and hydrophobic interactions. The simulated trajectories were then examined and visualized using VMD, and the amino acids within the distance of 5 Å from the enzyme were identified on SpyTag/ SpyCather for each snapshot. Given that these amino acids played an important role in the formation of the protein-protein interface between SpyTag/SpyCather and the enzyme, mutations in them could have a strong effect on the properties of the enzyme, including changes in the thermal elasticity of the enzyme. Here, we designed four types of mutations involving electrostatic interactions at the interface, and study the effect of these on the enzyme thermal resilience.

**Interactions at protein-protein interfaces**

The detailed chemo-physical contacts between enzyme and the SpyCather/SpyTag were automatically determined using the software of Ring[39] with the default parameters of the program. Based on these calculations, we observed the evolutions of the PPIs as the trajectories of dynamic system, thus characterizing the intermeidates states that might be important in inducing the docking of SpyRing to the enzyme，identifying changes in interactions located at protein-protein interfaces during the simulation, and highlighting key interactions determining the PPI. The interaction calculations were also used to analysis the perturbations due to mutations at the interface.

**Calculation of binding energy**

To estimate overall binding interaction between SpyRing and the enzyme and their perturbation due to mutations, free energy calculations were performed for selected systems. The binding free energy $\Delta G_{binding}$ of protein with ligand in solvent can be expressed as follows:

$$\Delta G_{binding} = G_{complex} - (G_{protein} + G_{ligand})$$

where $G_{complex}$ is the total binding energy of the protein-ligand complex, $G_{protein}$, $G_{ligand}$ stand for the respective binding energies of the protein and ligand, including van der Waals energy, electrostatic energy, polar solvation energy, and nonpolar solvation energy, etc. g_mmpbsa program[40, 41] was used to analyze the difference between binding energies of wild type and mutants. g_mmpbsa is developed from GROMACS and APBS. This tool computes the components of the binding energy using the MM-PBSA method, analyzed further using Python scripts, being able to obtain the final binding energy. Currently, g_mmpbsa does not include the calculation of an entropy term and therefore cannot give absolute binding energies. However, this tool is suitable for relative binding energy calculations comparing binding between different ligands and the same receptor.

Here, we considered the lichenase as a protein and the SpyRing structure (SpyCatcher/SpyTag) as the ligand. Considering that both parts were negatively charged, which created repulsion, and that there was a large amount of polar action inside the protein, we chose the amino acids located at the protein interface as objects for calculation.

In order to compare the SpyRing cyclization-based docking, we used the program ZDOCK[42] to investigate the random and free docking of SpyCatcher/SpyTag to lichenase without linking it to the end of the enzyme. ZDOCK gave 2000 random conformations, of which the one with the highest score was selected for further optimization using RosettaDock[43, 44]. Then, after the re-docking and optimizing of side chains by 10 cycles, the protein-protein interface energy was determined by Rosetta InterfaceAnalyzer[45]

## Results

***There is a stable interface between the enzyme and SpyCatcher/SpyTag***

For the initial cyclized lichenase complex, we performed three simulations to ensure the credibility of their results. After cyclization with tag/catcher, a significant interaction was introduced at the protein interface between the enzyme and the tag/catcher complex (Figure 1). Through MD simulation, we can see clearly: in the initial stage, the enzyme was separated from the tag/catcher, but after 20 ns, the two were closely combined. The simulation results showed that the centroid distance between enzyme and tag/catcher decreases from 4.8 nm to 3.2 nm, resulting in protein interface interaction, including electrostatic interaction, hydrogen bonding and hydrophobic interaction. Among the amino acids located at protein interfaces, a subset are conserved, including Asp30 near the N-terminus of the catcher, and Lys122, Asp124, located at the C-terminus of the catcher.

**Mutations altered the intermediate pathway of SpyRing docking with the enzyme**

*Mutation of neutral to positive charge* (T136K) *had no significant effect on thermal resileient of SpyRing* T136K showed two random results. In the initial stage of the simulation, Lys136 and Asp124 had strong electrostatic interaction, which made them close to each other. Later, they separated again which may due to the electrostatic repulsion of Lys122 to Lys136 (Figure 2a). However, if Lys136 was electrostatic repulsed by Lys122 at the initial stage, Asp124 cannot get close to it. This left enough space near Lys136. Asp49 and Lys136 had strong electrostatic interaction, which made lichenase twist to a certain extent (Figure 2b). But this did not have a clear impact on the cyclase overall stability, as manifested by the fact that the RMSD of the mutants was essentially comparable to that of the wild type. The Cα-Root mean square fluctuations showed that the mutations did not decrease the fluctuation of amino acid residues and even increased, indicating that the mutations did not confer a favorable effect on the stability of the catalytic site (Figure 3).

*Mutation of positive charge to negative charge (K122E) had no significant effect*

After the mutation of K122E, both Glu122 and Glu34/Glu35 had electrostatic interaction with Lys347. Glu122 and Asp124 formed the interaction with Lys347 first, Glu34 was also in close proximity to Lys347, but eventually left because of the electrostatic repulsive forces of the same charge (Figure S1a). Another possibility was that after Glu122 was in close proximity to Lys347 with Asp124, it caused them to leave because of the stronger electrostatic interaction of Glu35 with Lys347. At the same time, Glu35 was positioned next to Lys347 and the state was maintained (Figure S1b). However, the opposite direction of the charge mutation did not render the cyclase structure more stable, as indicated by the RMSD of the mutants largely consistent with the wild type. The Cα-Root mean square fluctuations showed that the mutations did not affect the fluctuation of most amino acid residues (Figure 4).

*Mutation of positive charge to neutral (K122G) reduced RMSD*   Lys348 with Asp124 and Glu35 would all result in electrostatic attraction. In different simulations, the preference of Lys348 was uncertain, and it tended to Asp124 or Glu35 randomly. Due to the mutation of K122G, the positive charge of Lys122 was eliminated, accompanied by the electrostatic repulsion with Lys348 and the elimination of the attraction with Glu35, making Lys348 and Glu35/Asp124 close together (Figure S2). This made the structure of the cyclase complex more stable and showed a reduced RMSD in the backbone dynamics. The Cα-Root mean square fluctuations showed that the mutations decreased the fluctuation of amino acid residues partially located at the active site, which indicated that the mutations played a certain role in the stabilization of the catalytic site (Figure 5).

*Mutation of negative charge to neutral (D124G) reduced RMSD*   After D124G mutation, the negative charge of Asp124 was eliminated and the electrostatic forces between Asp124 and Lys348 disappeared. At the same time, Lys348, Asp30, Glu34, Lys122 were electrostatically attractive to each other and all four maintained an equilibrium structure (Figure 6). Due to the enhanced interaction, the structure of cyclase complex was more stable, and the RMSD was reduced to some extent. The Cα-Root mean square fluctuations showed that the mutations did not affect the fluctuation of most amino acid residues (Figure 7).

**Mutations introduce new interactions at the interface**

Upon introduction of a mutation, interactions located at the protein interface produce certain changes. Upon addition of charge, we could find that a new electrostatic interaction (Lys136 with Asp124 / Asp49) was introduced at the protein interface, but this role might simply play a guiding effect, exist only as an intermediate process, and finally disappear. When mutating the charge to the opposite charge, there were two possibilities for the hydrogen bonding interaction of Lys122 / Glu122 with Lys347 to be retained or lost, with the chance of forming a new electrostatic interaction (Glu35 with Lys347). Upon reducing the charge, the original electrostatic interactions were eliminated, and the intermediate process had the potential to form new electrostatic interactions transiently (Asp30 / Glu34 with Lys347) but not eventually.

**SpyRing introduces a more stable interface than random docking**

The degree of negativity in the binding energy conferred by the different mutations all increased compared with the wild type, implying that the mutants are all more stable. The electrostatic energy was basically consistent with the trend of binding energy, which can be found to contribute a large part in the binding energy (Figure 8a). The most stable of these was K122G, which was in accordance with the results of RMSD. In the binding energy, T136K with K122E were both more stable than the wild-type, but showed no significant change in RMSD, which may be due to that we only selected the amino acids at the interface to calculate the binding energy, neglecting the electrostatic energy and the large amount of solvent polarity energy inside the protein, which also have some effect on our results.

However, when we randomly docked lichenase with spytag/spycatcher, we could find that the change of binding free energy of products after monomer mutation docking is subtle (Figure 8b). Therefore, compared with the binding energy difference before and after the cyclase mutation, it was obvious that the effect of interface mutation was significant, which indicated that the introduction of mutation at the interface would have a great impact on the stability of the cyclase complex.

In terms of structure, there was a significant difference between cyclase complex and random docking complex (Table S1). The former had a closer binding, and the protein-protein interface was larger, with more interface interaction, including van der Waals force, hydrogen bond, electrostatic interaction force and pi-pi stack; while the interface of random docking products was small and the interaction was less.

## Discussions

It was well recognized that the interaction at the interface takes precedence over Cys-Cys, opposite charge (electrostatic interaction) and hydrophobicity, while the interface residues are mostly characterized by hydrophobicity, aromaticity and long side chain[46, 47]. Therefore, we investigated the effect of changing the protein-protein interface interaction on the thermostability of the enzyme, by changing the electrostatic interactions.

Electrostatic force may play an electrostatic guiding role[48] in the process of protein recognition, which can promote the formation of protein-protein interaction and provide directivity[49]. Moreover, it can diffuse in a large range and improve the binding rate[50]. One of the unique features of SpyRing cyclization revealed by simulation calculations in this work is the presence of electrostatic interactions that play a guiding and directional role in the whole process of SpyCatcher/SpyTag docking with enzyme. These interactions did not have to survive to the end of the docking process, nor are they the most important forces maintaining protein-protein interfaces. But they can bring oppositely charged amino acids and their neighboring a.a. close to each other, which in turn promotes other interactions. This is consistent with previous studies[51]. From the calculation results of combined energy, it can be seen that electrostatic energy contributes a lot to the sum of combined energy. This proves that the contribution of electrostatic interaction between two molecules at 10 Å distance is higher than that of other energy components. Therefore, for protein interface, the change of electrostatic interaction may have a greater impact.

Increasing the temperature (from 333 K to 348 K) also had a significant effect on the protein-protein interface interaction and binding energy. From Table S2, we can see that the protein-protein interface decreased after heating, and the biggest feature was that the electrostatic interaction were reduced, because the increase of temperature would lead to the breaking of ionic bonds; however, the hydrogen bond and van der Waals force had no significant effect. In addition, from the point of view of interface energy, the van der Waals energy, electrostatic energy and apolar solvation energy decreased and the polar solvation energy increased with increasing temperature; the total binding energy was decreased (Table S3).

The present work introduces mutations at the protein interface to alter its interactions and thereby explores the effect on the thermal stability of the enzyme. We found that mutating critical amino acids at protein interfaces altered interface interactions, leading to diverse outcomes. In addition to alterations in electrostatic interactions, modifications to the hydrophobic interactions may also be considered. Apart from this, perhaps alterations can be made from the perspective of hotspot residues. We can make different adaptations of protein interfaces to different needs to ultimately serve a specific purpose.

# Conclusions

In summary, in this paper we took lichenase as an example to simulate and analyzed the dynamics of wild-type cyclase complex, and introduced three kinds of mutation: increase charge, mutation to opposite charge and decrease charge. By calculation, we found that the introduction of mutations at the protein-protein interface did have a significant effect on the interface interaction and energy, thus affecting the thermal stability of the enzyme. Appropriate mutation will produce new interfacial interaction, improve the binding energy between enzyme and cyclization structure, thus improve the thermal stability of enzyme. Our results might provide a reasonable basis for the new structure design of SpyRing cyclization reaction.

**Acknowledgements**

This work was supported, in part, by the National Key Research and Development Program of China (Grant No. 2019YFA0905700, 2017YFC1600900).

Figures & Tables:

Figure 1. Protein structure of wild type before and after simulation.

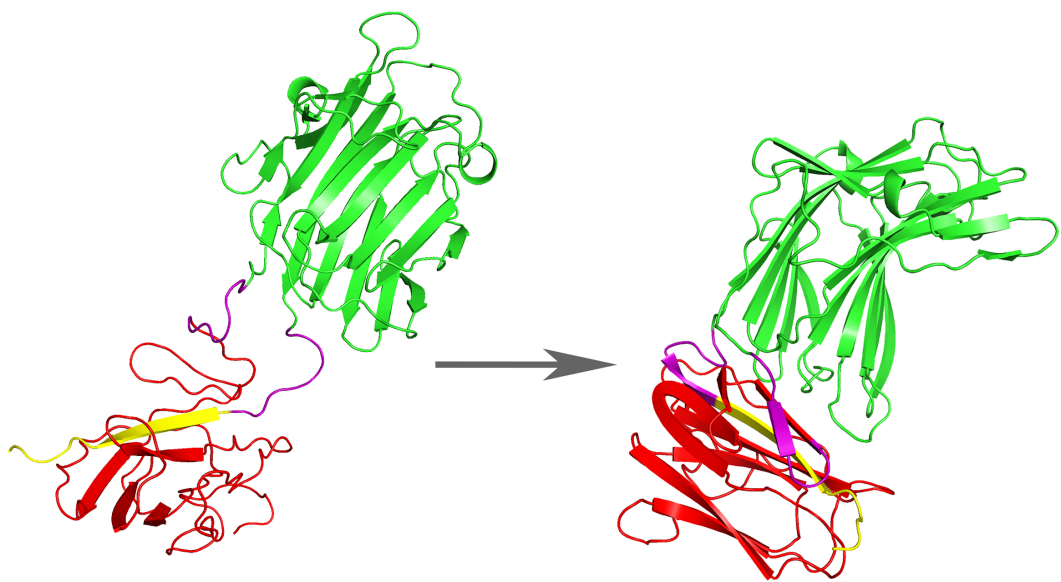

Figure 2. Structure in simulated trajectory of T136K. (a) and (b) are two simulation results. (a) Lys136 and Asp124 were close to each other first, and then separated again, which might be due to the electrostatic repulsion of Lys122. (b) Instead of Asp124, Asp49 and Lys136 had strong electrostatic interaction and were close to each other.

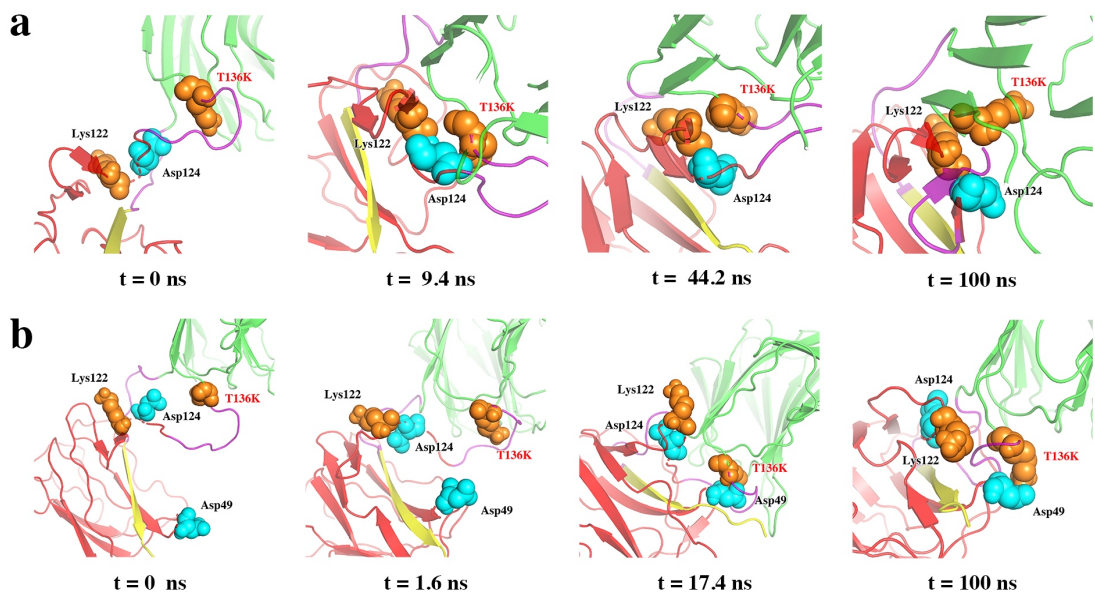

Figure 3. RMSD and RMSF (the inset of Figure 3) of lichenase with the mutation of T136K. The red arrows in the RMSF diagram refer to the active sites.

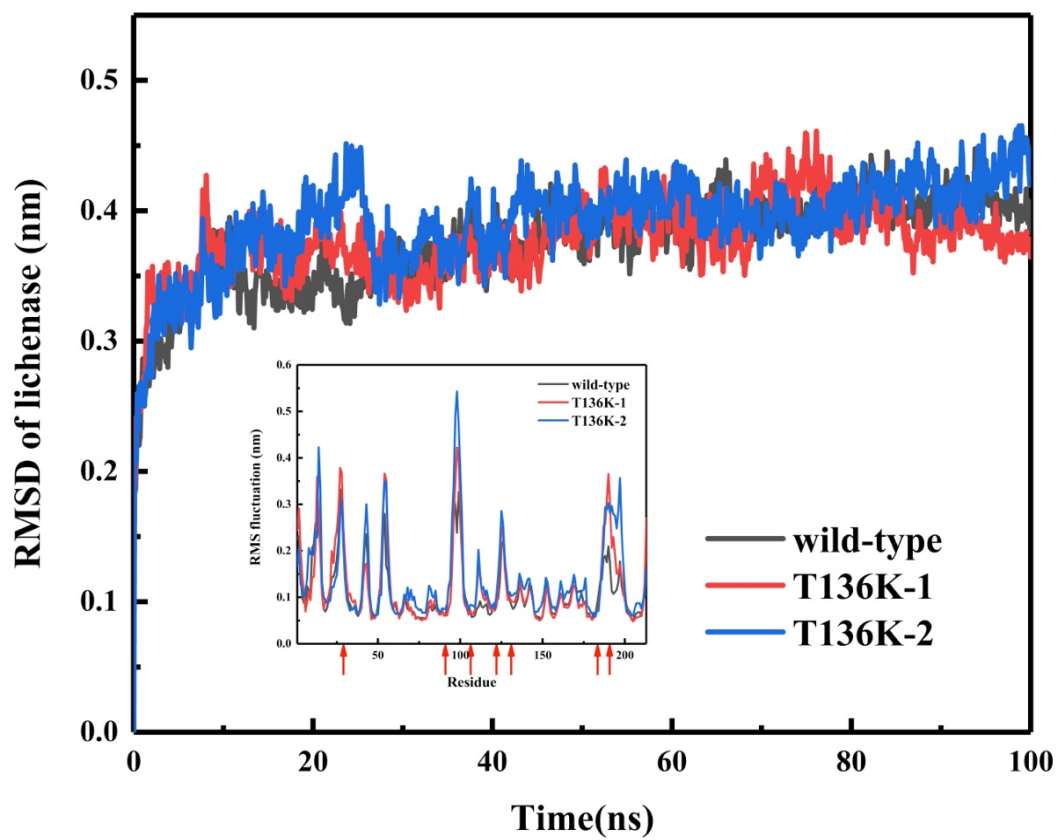

Figure 4. RMSD and RMSF (the inset of Figure 4) of lichenase with the mutation of K122E. The red arrows in the RMSF diagram refer to the active sites.

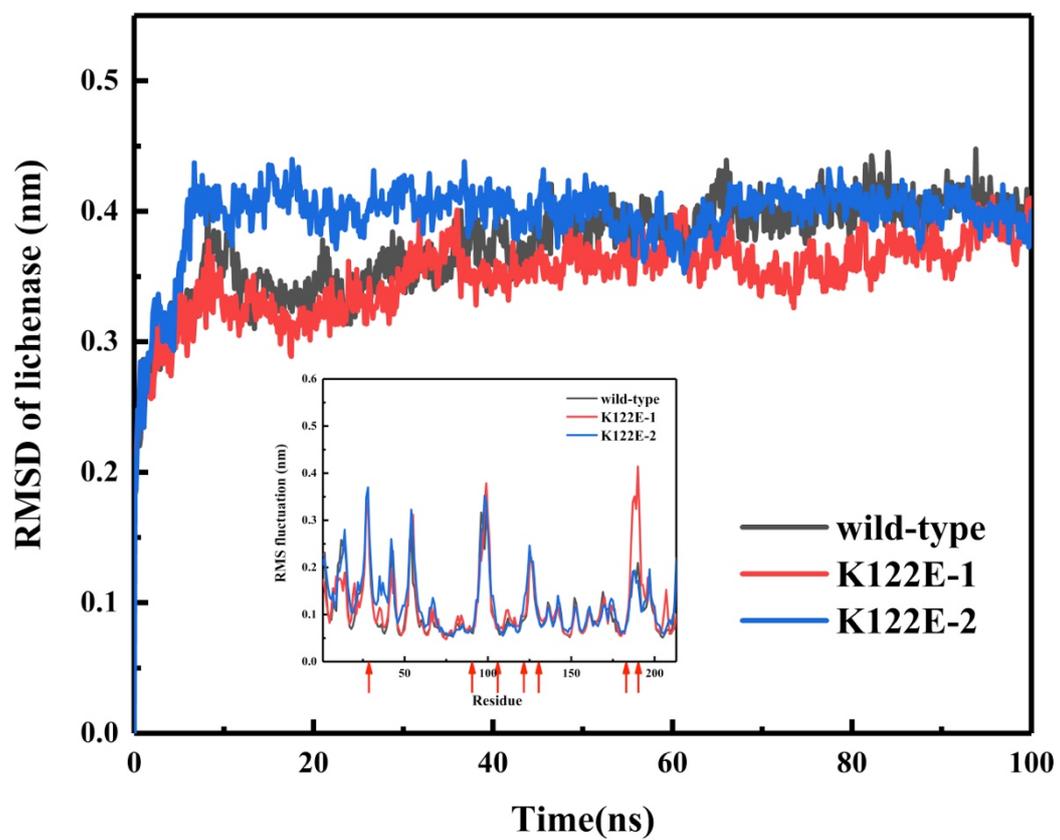

Figure 5. RMSD and RMSF (the inset of Figure 5) of lichenase with the mutation of K122G. The red arrows in the RMSF diagram refer to the active sites.

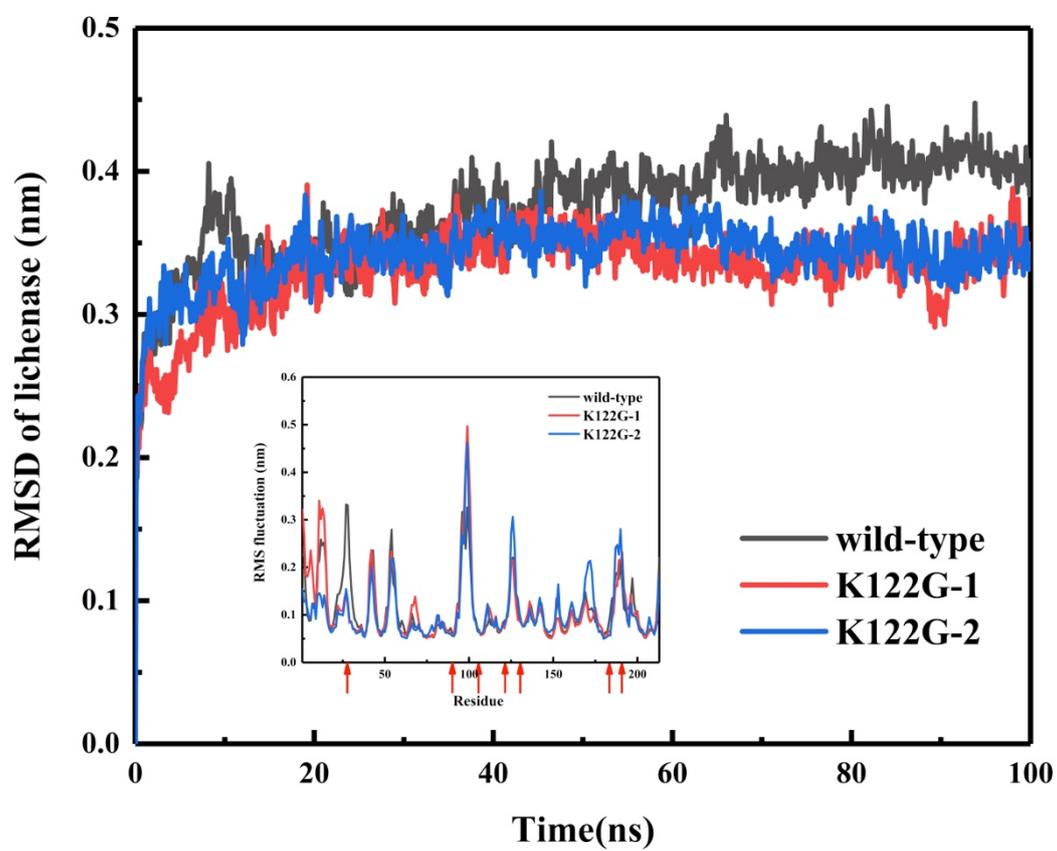

Figure 6. Structure in simulated trajectory of D124G. (a) and (b) are two similar simulation results. Lys348, Asp30, Glu34, Lys122 were electrostatically attractive to each other and all four maintained an equilibrium structure.

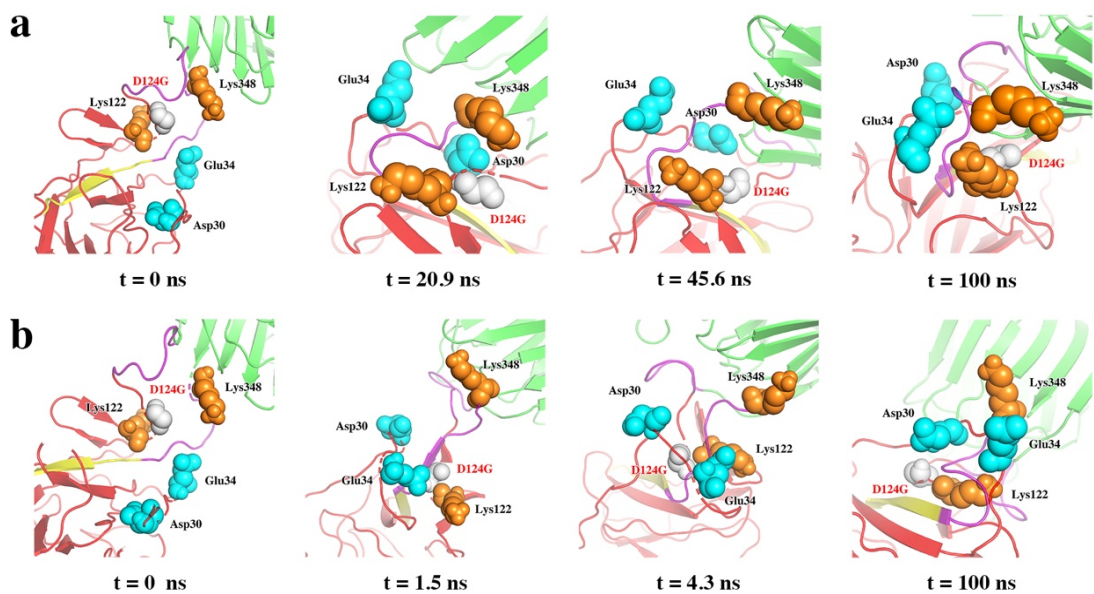

Figure 7. RMSD and RMSF (the inset of Figure 7) of lichenase with the mutation of D124G. The red arrows in the RMSF diagram refer to the active sites.

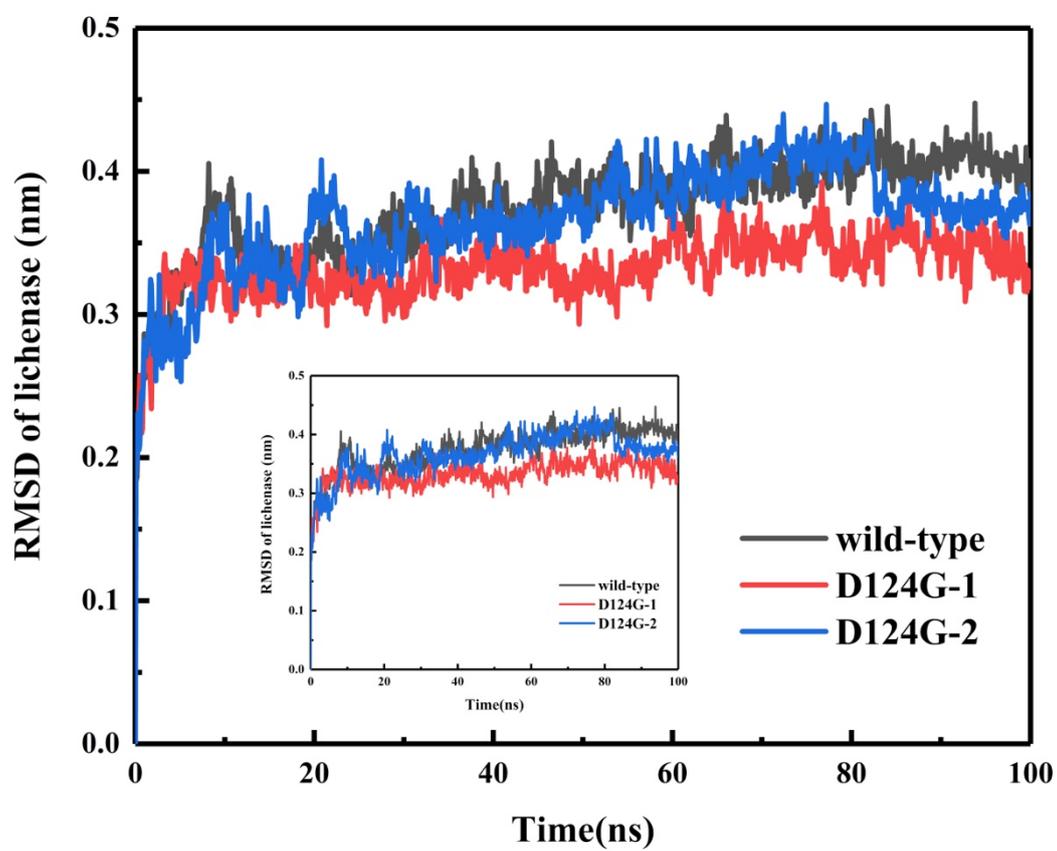

Figure 8. (a) Binding energy and electrostatic energy of wild type and mutants. (b) Binding energy of complexes formed by random docking. REU means Rosetta energy unit.

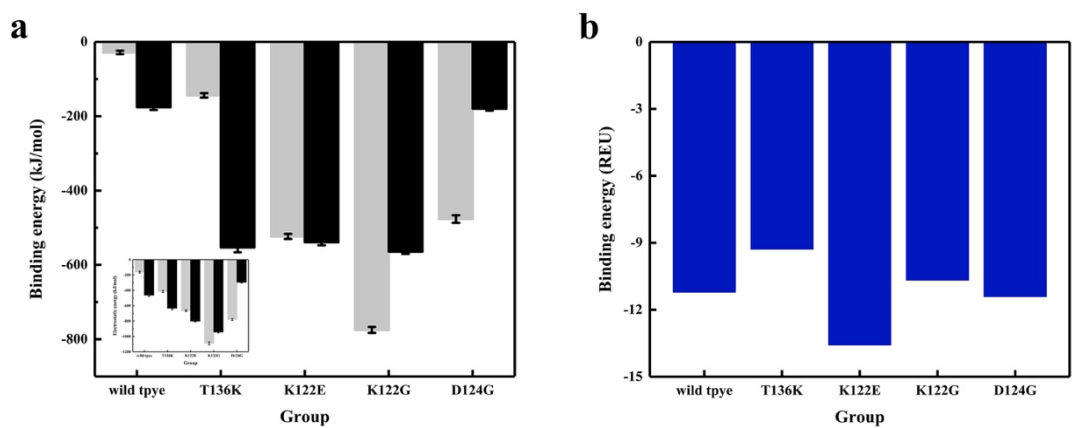

Figure S1. Structure in simulated trajectory of K122E. (a) and (b) are two simulation results. (a) Glu122 and Asp124 formed the interaction with Lys347 first and maintained until the end; Glu34 was also in close proximity to Lys347, but eventually left. (b) Glu122 and Asp124 were closed to Lys347, and then left. Instead, Glu35 was next to Lys347 and remained in this state.

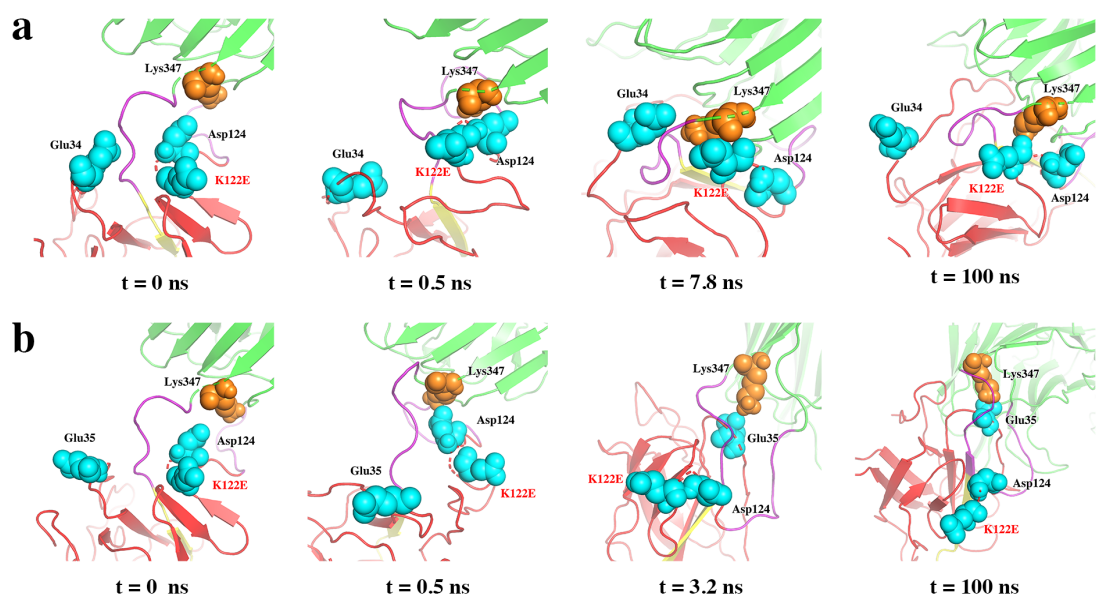

Figure S2. Structure in simulated trajectory of K122G. (a) and (b) are two simulation results. (a) Lys348 preferred Glu35 to Asp124 in electrostatic attraction. (b) Lys348 preferred Asp124 to Glu35 in electrostatic attraction.

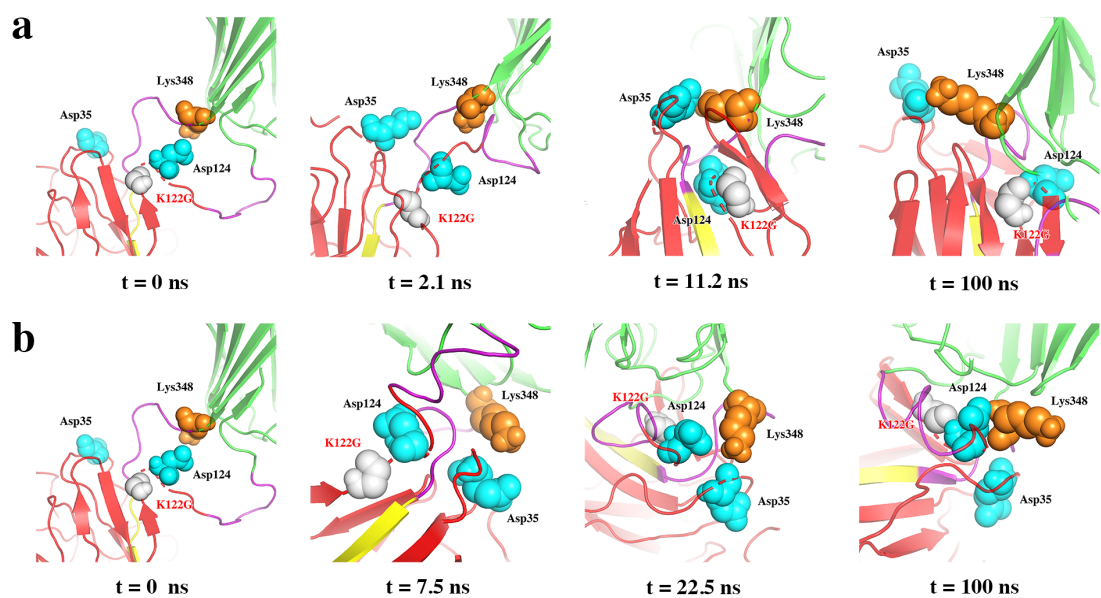

Table S1: The difference of interface between cyclase complex for simulation and random docking products. The amino acid sequence numbers are used according to the simulated cyclase complex.

| Group | | Tag/catcher | lichenase | Interaction |
|---|---|---|---|---|
| For simulation | Wild type | Asp30 | Asn288 | VDW |
| | | Met31 | Asn288 | VDW |
| | | Met31 | Asn304 | VDW |
| | | Thr32 | Asn288 | Hbond |
| | | Glu34 | Lys291 | Hbond |
| | | Lys42 | Asn288 | Hbond |
| | | Arg46 | Thr235 | VDW |
| | | Lys122 | Lys137 | Hbond |
| | | Lys122 | Lys347 | Hbond |
| | | Asp124 | Gly205 | Hbond |
| | | Asp124 | Tyr206 | Hbond/VDW |
| | | Asp124 | Pro310 | VDW |
| | | Asp124 | Lys347 | Ionic/Hbond |
| | | His126 | Pro231 | VDW |
| | | His126 | Thr309 | VDW |
| | | His126 | Thr322 | Hbond |
| | | His358 | Tyr206 | Pipistack/VDW |
| | | His358 | Pro287 | VDW |
| | | Val362 | Thr235 | VDW |
| For random docking | Wild type | Asp124 | Trp237 | VDW |
| | | His126 | Trp237 | VDW |
| | T136K | Gly118 | Pro231 | VDW |
| | | Lys122 | Gln305 | VDW |
| | | His126 | Gly262 | Hbond |
| | | His126 | Asn263 | VDW |
| | K122E | Lys119 | Asn259 | VDW |
| | | Ala120 | Ans259 | VDW |
| | K122G | Gln111 | Gln305 | Hbond |
| | | Ala120 | Asn259 | VDW |
| | D124G | Gly118 | Gly322 | Hbond |
| | | Lys119 | Asn162 | VDW |
| | | Ala120 | Thr321 | VDW |

Table S2: The change of interface with increasing temperature (from 333 K to 348 K).

|  | Tag/catcher | lichenase | Interaction |
|---|---|---|---|
| Wild type (333 K) | Asp30 | Asn288 | VDW |
|  | Met31 | Asn288 | VDW |
|  | Met31 | Asn304 | VDW |
|  | Thr32 | Asn288 | Hbond |
|  | Glu34 | Lys291 | Hbond |
|  | Lys42 | Asn288 | Hbond |
|  | Arg46 | Thr235 | VDW |
|  | Lys122 | Lys137 | Hbond |
|  | Lys122 | Lys347 | Hbond |
|  | Asp124 | Gly205 | Hbond |
|  | Asp124 | Tyr206 | Hbond/VDW |
|  | Asp124 | Pro310 | VDW |
|  | Asp124 | Lys347 | Ionic/Hbond |
|  | His126 | Pro231 | VDW |
|  | His126 | Thr309 | VDW |
|  | His126 | Thr322 | Hbond |
|  | His358 | Tyr206 | Pipistack/VDW |
|  | His358 | Pro287 | VDW |
|  | Val362 | Thr235 | VDW |
| Wild type (348 K) | Gln25 | Thr235 | VDW |
|  | Gln27 | Pro287 | Hbond |
|  | Gly29 | Gln286 | VDW |
|  | Gly29 | Asn288 | Hbond |
|  | Asp30 | Gln286 | Hbond/VDW |
|  | Arg46 | Gly234 | Hbond |
|  | Gly50 | Thr235 | Hbond |
|  | Lys51 | Thr235 | Hbond |
|  | Lys51 | Trp237 | VDW |
|  | Asp124 | Tyr206 | Hbond |
|  | Asp124 | Lys347 | Ionic/Hbond |
|  | His126 | Asp233 | Ionic/Hbond |
|  | His358 | Tyr206 | Pipistack/VDW |
|  | His358 | Asp233 | Hbond |
| K122G (333 K) | Gln113 | Ser139 | Hbond |
|  | Gly118 | Gln202 | VDW |
|  | Ala120 | Gly137 | VDW |
|  | Gly122 | Gly137 | Hbond |
|  | Gly122 | Lys347 | Hbond/VDW |
|  | Asp124 | Lys347 | Ionic/Hbond/VDW |
|  | His126 | Tyr206 | Pipistack/VDW |
|  | His126 | Tyr308 | Hbond |
| K122G (348 K) | Gly122 | Lys347 | Hbond |

Table S3: The change of each component in binding energy with increasing temperature (from 333 K to 348 K). The unit of energy is kJ/mol.

|  | Van der Waal | Electrostatic | Polar solvation | Apolar solvation | Binding |
|---|---|---|---|---|---|
| Wild type (333 K) | -36.6±2.0 | -462.0±16.8 | 333.4±12.4 | -9.0±0.3 | -175.5±7.7 |
| Wild type (348 K) | -86.1±3.1 | -164.6±20.1 | 203.1±15.7 | -15.0±0.6 | -61.6±9.2 |
| K122G (333 K) | -44.2±1.7 | -941.7±12.9 | 433.4±9.2 | -11.7±0.1 | -564.45±5.6 |
| K122G (348 K) | -8.4±0.8 | -290.7±11.7 | 100.5±9.0 | -3.0±0.2 | -202.1±3.4 |